\PassOptionsToPackage{unicode}{hyperref}
\PassOptionsToPackage{hyphens}{url}
\PassOptionsToPackage{dvipsnames,svgnames,x11names}{xcolor}
\PassOptionsToPackage{page,toc,title}{appendix}
\PassOptionsToPackage{normalem}{ulem}

\documentclass[]{article}

\usepackage[T1]{fontenc}
\usepackage[utf8]{inputenc}
\usepackage{lmodern}
\usepackage{iftex}
\ifPDFTeX
  \usepackage{textcomp}
\else
  \usepackage{unicode-math}
  \defaultfontfeatures{Scale=MatchLowercase}
  \defaultfontfeatures[\rmfamily]{Ligatures=TeX,Scale=1}
\fi

\usepackage{xcolor}
\usepackage[top=30mm,left=30mm,heightrounded]{geometry}
\usepackage{graphicx}
\usepackage{adjustbox}
\usepackage{wrapfig}
\usepackage{pdflscape}
\usepackage{tcolorbox}
\usepackage{float}
\usepackage{caption}
\usepackage{subcaption}

\usepackage{array}
\usepackage{booktabs}
\usepackage{tabularx}
\usepackage{tabulary}
\usepackage{longtable}
\usepackage{multirow}
\usepackage{makecell}
\usepackage{threeparttable}
\usepackage{threeparttablex}
\usepackage{colortbl}

\newcolumntype{L}[1]{>{\raggedright\arraybackslash}p{#1}}

\newlength{\colwidth}
\usepackage{etoolbox}
\usepackage{listings}
\usepackage{xpatch}

\usepackage{xurl}
\usepackage{hyperref}
\usepackage{bookmark}
\hypersetup{
 pdftitle={What do people expect from Artificial Intelligence},
  colorlinks=true,
  linkcolor={blue},
  filecolor={Maroon},
  citecolor={Blue},
  urlcolor={Blue},
  pdfcreator={LaTeX}
}

\usepackage{appendix}
\usepackage{ulem}
\usepackage{authblk}
\usepackage{orcidlink}

\makeatletter
\patchcmd\longtable{\par}{\if@noskipsec\mbox{}\fi\par}{}{}
\makeatother
\IfFileExists{footnotehyper.sty}{\usepackage{footnotehyper}}{\usepackage{footnote}}
\makesavenoteenv{longtable}

\makeatletter
\ifx\paragraph\undefined\else
  \let\oldparagraph\paragraph
  \renewcommand{\paragraph}[1]{\oldparagraph{#1}\mbox{}}
\fi
\ifx\subparagraph\undefined\else
  \let\oldsubparagraph\subparagraph
  \renewcommand{\subparagraph}[1]{\oldsubparagraph{#1}\mbox{}}
\fi
\makeatother

\makeatletter
\@ifundefined{KOMAClassName}{%
  \IfFileExists{parskip.sty}{%
    \usepackage{parskip}
  }{%
    \setlength{\parindent}{0pt}
    \setlength{\parskip}{6pt plus 2pt minus 1pt}}
}{%
  \KOMAoptions{parskip=half}}
\makeatother

\newlength{\cslhangindent}
\newlength{\csllabelwidth}
 {\begin{list}{}{%
  \setlength{\itemindent}{0pt}
  \setlength{\leftmargin}{0pt}
  \setlength{\parsep}{0pt}
  \ifodd #1
   \setlength{\leftmargin}{\cslhangindent}
   \setlength{\itemindent}{-1\cslhangindent}
  \fi
  \setlength{\itemsep}{#2\baselineskip}}}
 {\end{list}}

\makeatletter
\let\@cite@ofmt\@firstofone
\def\@biblabel#1{}
\def\@cite#1#2{{#1\if@tempswa , #2\fi}}
\makeatother

\makeatletter
\@ifundefined{c@chapter}
  {\newfloat{codelisting}{h}{lop}}
  {\newfloat{codelisting}{h}{lop}[chapter]}
\floatname{codelisting}{Listing}

\makeatother

\providecommand{\tightlist}{%
  \setlength{\itemsep}{0pt}\setlength{\parskip}{0pt}}

\title{{What do people expect from Artificial Intelligence? Public opinion on alignment in AI moderation from Germany and the United States}}
\author[1]{Andreas Jungherr \orcidlink{0000-0003-2598-2453}}
\author[2]{Adrian Rauchfleisch \orcidlink{0000-0003-1232-083X}}
\affil[1]{University of Bamberg}
\affil[2]{National Taiwan University}
\date{\today}

\begin{document}

\maketitle
\begin{abstract}
Recent advances in generative Artificial Intelligence have raised public awareness, shaping
expectations and concerns about their societal implications. Central to
these debates is the question of AI alignment---how well AI systems meet
public expectations regarding safety, fairness, and social values.
However, little is known about what people expect from AI-enabled
systems and how these expectations differ across national contexts. We
present evidence from two surveys of public preferences for key
functional features of AI-enabled systems in Germany (n = 1800) and the
United States (n = 1756). We examine support for four types of alignment
in AI moderation: accuracy and reliability, safety, bias mitigation, and
the promotion of aspirational imaginaries. U.S. respondents report
significantly higher AI use and consistently greater support for all
alignment features, reflecting broader technological openness and higher
societal involvement with AI. In both countries, accuracy and safety
enjoy the strongest support, while more normatively charged goals---like
fairness and aspirational imaginaries---receive more cautious backing,
particularly in Germany. We also explore how individual experience with
AI, attitudes toward free speech, political ideology, partisan
affiliation, and gender shape these preferences. AI use and free speech
support explain more variation in Germany. In contrast, U.S. responses
show greater attitudinal uniformity, suggesting that higher exposure to
AI may consolidate public expectations. These findings contribute to
debates on AI governance and cross-national variation in public
preferences. More broadly, our study demonstrates the value of
empirically grounding AI alignment debates in public attitudes and of
explicitly developing normatively grounded expectations into theoretical
and policy discussions on the governance of AI-generated content.
\end{abstract}


\section{Introduction}\label{introduction}

Recent successes of AI-enabled systems\footnote{We define AI-enabled
  systems as technologies or services that apply artificial intelligence
  methods---such as machine learning, deep learning, and generative
  modeling--- to perform tasks intelligently. Intelligence is understood
  here as ``as that quality that enables an entity to function
  appropriately and with foresight in its environment.'' (Nilsson, 2009,
  p.~xiii). These systems are capable of learning from data, recognizing
  patterns, making context-aware decisions, and generating
  content---often autonomously or with minimal human input---across a
  range of applications including natural language processing, computer
  vision, and robotics.} such as \emph{ChatGPT}, \emph{DALL·E}, and
\emph{Midjourney} have significantly raised public awareness of
generative AI. These systems, and others like them, have made generative
models more accessible and user-friendly, resulting in widespread usage
and visibility. As a consequence, both the capabilities and limitations
of these technologies have entered public discourse, fueling growing
expectations as well as mounting concerns about their societal impact.

Among the key concerns emerging in this debate is the issue of AI
alignment---that is, the extent to which AI-enabled systems act in
accordance with the intentions and expectations of their developers
(Amodei et al., 2016; Bommasani et al., 2022; Gabriel, 2020; Hendrycks
et al., 2022). But despite the growing relevance of AI alignment, public
attitudes toward features of AI-enabled systems remain underexplored.
Yet such attitudes matter. Public opinion on digital governance often
reflects---though not always straightforwardly---partisan affiliations
and deeper ideological commitments (Jang et al., 2024; Rauchfleisch et
al., 2025; Rauchfleisch \& Jungherr, 2024). This is likely to hold true
for AI governance as well, although empirical insights remain scarce,
especially across countries.

This article addresses this research gap by presenting a comparative
study of public preferences for specific AI features: accuracy and
reliability, safety, bias mitigation, and the promotion of aspirational
imaginaries. We selected these features for their relevance in the
larger AI alignment debate. We compare survey responses from Germany (n
= 1800) and the United States (n = 1756) and examine the role of key
factors that may shape governance preferences on the individual, group,
and systems levels.

We find that respondents in Germany and the United States differ
significantly in their use of AI, with U.S. participants reporting
higher usage levels. These patterns reflect broader differences in
technological adoption between the two countries. When evaluating
preferences for AI moderation, both German and U.S. participants
strongly support features aiming for accuracy and reliability, as well
as safety. However, support declines for interventions aimed at
mitigating bias or promoting aspirational imaginaries, especially in
Germany. U.S. respondents consistently show stronger support across all
categories, consistent with their higher societal involvement with AI.
Experience with AI, political ideology, gender, and free speech
attitudes all shape individual preferences, though their associations
vary by country. In Germany, both AI experience and free speech
attitudes are stronger predictors of support than in the U.S., where
attitudes appear more uniformly developed. Political ideology is more
influential in the U.S. when it comes to aspirational portrayals. These
findings suggest that in contexts with lower AI exposure,
individual-level factors play a greater role in shaping attitudes, while
in high-exposure contexts like the U.S., public views on AI moderation
are more consolidated.

We highlight the value of clearly identifying distinct, normatively
grounded principles for the moderation of AI outputs and situating these
principles within broader theoretical debates about the role of speech
and culture. Moreover, we underscore the importance of systematically
eliciting public views on the normative foundations of AI interventions.
Anchoring discussions of AI governance in established theoretical
frameworks---while also refining these frameworks where necessary---and
seriously engaging with public attitudes are essential steps toward
advancing academic inquiry, informing practical implementation, and
enriching public discourse on AI governance. This paper contributes to
this broader development by examining AI alignment through the lens of
model output moderation.

\section{Explaining public preferences for features of AI-enabled
systems}\label{explaining-public-preferences-for-features-of-ai-enabled-systems}

\subsection{AI alignment}\label{ai-alignment}

As AI-enabled systems move from research contexts into widespread
professional and public use, people's preferences for how these systems
are governed become increasingly important for the success of AI models,
governance frameworks, and broader public trust or skepticism toward the
growing integration of AI into society. This introduces an attitudinal
and psychological dimension to the broader debate on AI alignment
(Amodei et al., 2016; Bommasani et al., 2022; Hendrycks et al., 2022).

AI alignment broadly refers to the challenge of ensuring that AI models
act in accordance with the intentions and values of their developers or
deployers, rather than deviating from them in harmful or unintended ways
(Gabriel, 2020; Ji et al., 2023; Leike et al., 2018; Ngo et al., 2022).
This is challenging enough in settings where goals can be clearly
specified, but alignment becomes even more challenging and contested
when extended to the expectations of the broader public---especially
given the significant variation in public attitudes, levels of
expertise, and degrees of engagement (Achintalwar et al., 2024; Padhi et
al., 2024). A key source of divergence is the debate over how much
control developers should exert over model outputs and for what purposes
that control should be applied.

\subsection{Preferences in the use of AI-enabled systems: Accuracy and
reliability, safety, bias mitigation, and promotion of aspirational
imaginaries}\label{preferences-in-the-use-of-ai-enabled-systems-accuracy-and-reliability-safety-bias-mitigation-and-promotion-of-aspirational-imaginaries}

\subsubsection{Accuracy and reliability}\label{accuracy-and-reliability}

People vary in their expectations toward AI and the preferences they
hold in using AI-enabled systems. On a foundational level, one important
characteristic in the workings of AI-enabled systems is accuracy. In
other words, AI models should be bounded by facts, albeit facts found in
data (Fourcade \& Healy, 2024; Hand, 2004; Smith, 2019).

Much of public debate around generative AI centers on the risks
associated with factually incorrect or misleading outputs---whether
these stem from flawed training data, faulty inference, or so-called
``hallucinations,'' in which models generate information that appears
authoritative but is entirely fabricated or inaccurate (Maleki et al.,
2024; Maynez et al., 2020). This is particularly problematic in domains
such as search, information retrieval, public discourse, education,
democratic participation, and professional usage contexts where users
rely on AI for accurate and trustworthy information (Jungherr, 2023;
Jungherr et al., 2024; Jungherr \& Schroeder, 2023; Spitale et al.,
2023; Weidinger et al., 2022). Generative models can, for example,
fabricate historical events, misattribute facts, disseminate false
claims about individuals, or simply return wrong responses to factual or
analytical queries. While these outputs often appear plausible, their
misleading or fictional nature can result in real-world harms.

Having people look for accuracy and reliability in the workings of
AI-enabled systems appears to be a natural precondition for their
widespread and systematic use. At the same time, there are good reasons
for preferring the moderation of AI-enabled systems that adjust purely
data-driven outputs.

\subsubsection{Safety}\label{safety}

We expect people to voice a preference for AI-enabled systems that take
harm prevention and safety seriously and moderate the outputs of their
models accordingly. Concerns about preventing \emph{harm} and ensuring
\emph{safety} often justify intentional deviations from strictly
data-driven model outputs (Anwar et al., 2024; Askell et al., 2021; Hao
et al., 2023; Hendrycks et al., 2022; Phuong et al., 2024). This
includes filtering outputs that could pose direct risks---such as
generating instructions for building weapons, carrying out terrorist
attacks, or engineering pathogens. However, safety concerns also extend
to outputs that violate individual rights, such as privacy breaches or
unauthorized use of intellectual property. In these instances,
moderating or restricting factual content based on model-learned
patterns becomes a necessary intervention.

Safety-oriented moderation is not a new concept in digital governance.
Public support for such moderation is well-documented in debates
surrounding digital speech. While there is variation between countries
in what is perceived as severely harmful content (Jiang et al., 2021),
there is general support to moderate harmful content on online platforms
(Kozyreva et al., 2023; Pradel et al., 2024). Thus, the safety-focused
moderation of AI systems might appear to be a comparable case. However,
the impersonal and machine-generated nature of AI content may introduce
new dynamics in how people perceive and evaluate such moderation. These
differences could lead to different explanatory factors shaping public
attitudes and preferences in the context of AI.

\subsubsection{Bias mitigation}\label{bias-mitigation}

Another feature people might be looking for in their preferences for
AI-enabled systems is the \emph{mitigation of bias} to promote
\emph{fairness} (Askell et al., 2021; Barocas et al., 2023; Hao et al.,
2023). This is an intervention specifically made necessary due to the
workings of AI and, therefore, has no direct equivalent in the
discussion of speech moderation in other digital contexts. AI models do
not access objective facts about the world directly, but only
representations of those facts as encoded in data (Fourcade \& Healy, 2024; Hand,
2004; Smith, 2019). As a result, they are prone to reproducing the
biases embedded in their training data (Bianchi et al., 2023; Friedrich,
Brack, et al., 2024; Friedrich, Hämmerl, et al., 2024; Hofmann et al.,
2024; Tao et al., 2024; Weidinger et al., 2021). When AI systems are
bound to such biased representations, their outputs may reinforce
harmful distortions or inequalities.

Bias, in this context, can be understood as a divergence between the
distribution of a variable in the model's output---or training
data---and its distribution in the real world. Moderation aimed at
fairness seeks to adjust these outputs so that they better reflect a
more accurate or equitable distribution rather than the skewed patterns
found in data. This may involve correcting underrepresentation,
countering stereotypes, or highlighting marginalized perspectives.

While such interventions may seem normatively desirable, they are often
politically and ethically contested (Binns, 2018; Gabriel, 2020).
Efforts to mitigate bias inevitably raise questions about what
constitutes a fair or accurate representation of the world---and who
gets to decide which absences or distortions in data should be
corrected. In this sense, fairness-focused moderation confronts not only
technical challenges but also deeper disagreements about knowledge,
representation, and justice. This will likely shape who supports or
actively demands these interventions in AI-enabled systems.

\subsubsection{Promoting aspirational
imaginaries}\label{promoting-aspirational-imaginaries}

Another reason to adjust the output of AI models is also driven by
specific characteristics and workings of AI. This is less about
correcting inherent risks or drawbacks of AI, but instead about
capitalizing on AI's inherently promising opportunities for societal
change for the better. There is a normative argument that generative AI
should not simply reflect the world \emph{as it is}, but instead
contribute to envisioning the world as \emph{it should be}. In this
view, moderation is not about aligning outputs strictly with factual
distributions, but with aspirational ideals. Rather than being bound by
present realities, AI outputs would be shaped by values that seek to
challenge and transform those realities.

This argument draws on long-standing traditions in social theory,
cultural criticism, and pragmatist philosophy, which emphasize the role
of narrative, discourse, and imagination in promoting social justice,
change, and solidarity (Benjamin, 2025; Dewey, 1916, 1934; Rorty, 1989).
Culture and discourse, including the outputs of generative AI, need not
reproduce historical inequalities, exclusions, or injustices. Instead,
they can offer alternative imaginaries---visions of more inclusive
futures that expand the moral and social boundaries of the ``we,''
making visible those historically marginalized or silenced.

Moderation along these lines could involve deliberately shifting the
distribution of model outputs---even when those outputs align with
real-world data---if that distribution is understood to reflect
structural inequality or discrimination. Generative AI might thus serve
as a cultural and political tool, one that fosters empathy, reimagines
collective futures, and contributes to the extension of solidarity.

However, such interventions are particularly vulnerable to public
contestation. By explicitly using AI to pursue social or political
goals, they raise fundamental questions about legitimacy: Who decides
what the world should look like? Which visions of justice or inclusion
are privileged? And to what extent should AI systems be used as
instruments of normative transformation? Accordingly, we expect the
promotion of aspirational imaginaries not to be a universally held
preference for the working of AI-enabled systems.

\subsubsection{Preferences for features of AI-enabled
systems}\label{preferences-for-features-of-ai-enabled-systems}

These observations point to likely differences in the preferences for
approaches guiding the adjustment of outputs of AI-enabled systems.
Specifically, attitudes are likely to vary depending on the underlying
rationale. Moderation aimed at ensuring accuracy and reliability, or
safety---such as preventing harm, illegal activity, or violations of
individual rights---is likely to enjoy broad public support, as these
goals align with widely accepted norms and relatively uncontroversial
forms of risk prevention. In contrast, interventions designed to
mitigate bias or promote aspirational imaginaries introduce greater
ambiguity, normative complexity, and potential for political
disagreement. Bias mitigation involves contested judgments about what
constitutes fairness or representational accuracy (Binns, 2018; Gabriel,
2020), while aspirational goals go further by seeking to reshape
cultural narratives or advance particular visions of a better future.
Such aims can trigger concerns about legitimacy, overreach, and
ideological bias. Accordingly, the motivation behind a given moderation
decision is likely to influence how it is received by the public, with
support declining as the rationale shifts from widely shared safety
concerns to more contested and value-laden objectives.

\begin{tcolorbox}[colback=gray!5!white, colframe=gray!80!black, title=Research Question 1]
To what extent do public preferences vary between different
principles of AI governance, namely accuracy and reliability, safety,
the mitigation of bias, or the promotion of aspirational societal
values?
\end{tcolorbox}

\subsection{Explaining preferences: The role of
involvement}\label{explaining-preferences-the-role-of-involvement}

We propose a model in which preferences regarding the adjustment of
AI-generated outputs are shaped by varying degrees of personal and
collective involvement. This involvement can manifest on multiple
levels:

\begin{itemize}
\tightlist
\item
  At the individual level, it may reflect personal experiences with AI
  and relevant related attitudes.\\
\item
  At the group level, it may be influenced by membership in a group that
  is particularly affected by or sensitive to AI-based interventions.\\
\item
  At the systemic level, it may depend on whether an individual resides
  in a country with a strong or weak technological infrastructure and
  relationship to digital innovation.
\end{itemize}

In the following, we elaborate on the rationale behind each of these
dimensions.

\subsubsection{Individual-level involvement: Experience \&
Values}\label{individual-level-involvement-experience-values}

Individual support for AI content moderation is shaped not only by how
frequently people use AI technologies but also by the values and beliefs
they bring to evaluating their use. We distinguish two dimensions of
individual-level involvement with AI: experiential and ideational.

\paragraph{\texorpdfstring{\emph{Experiential
Involvement}}{Experiential Involvement}}\label{experiential-involvement}

Personal involvement with AI can take different forms. One case involves
individuals who actively use AI for personal or professional reasons.
These users experience the technology firsthand and can develop more
elaborate and specific preferences regarding its regulation than those
who do not (Horowitz et al., 2024; Horowitz \& Kahn, 2021). We expect
this familiarity to translate into differentiated preferences toward AI
moderation.

People with limited exposure to or interaction with AI-enabled systems
may approach them with greater skepticism or uncertainty. This
skepticism may lead to a heightened demand for external safeguards,
particularly those framed around safety concerns. Moderation aimed at
reducing bias or promoting aspirational portrayals, however, may appear
unnecessary or overly intrusive to individuals with low AI involvement
since these interventions presuppose familiarity with how AI systems
operate.

In contrast, individuals with greater hands-on experience using
AI-enabled systems may have more specific views on the systems'
capabilities and limitations. This familiarity may foster a greater
appreciation for moderation tasks that are specific to AI-generated
content---particularly interventions aimed at bias mitigation or the
promotion of aspirational social values.

\paragraph{\texorpdfstring{\emph{Ideational
Involvement}}{Ideational Involvement}}\label{ideational-involvement}

In addition to usage-based experience, individual attitudes toward AI
moderation are shaped by broader normative commitments and political
values. AI moderation can be seen as a special case of speech governance
more broadly (Dabhoiwala, 2025; Kosseff, 2023; Mchangama, 2022). As
such, individual support for AI interventions is likely to reflect how
people understand and prioritize free expression.

People who view free speech as a foundational democratic right may
oppose moderation efforts---especially those perceived as ideological or
normative in nature. In contrast, those who understand speech as
something that can and should be regulated in the interest of societal
fairness or safety may be more supportive of AI content moderation
(Rauchfleisch \& Jungherr, 2024; Riedl et al., 2021, 2022).

We therefore expect individuals who strongly support free speech to be
more critical of AI moderation interventions aimed at shaping content in
terms of fairness or aspirational values, while potentially supporting
moderation grounded in factual accuracy or safety.

Beyond attitudes toward free speech, broader political
ideology---particularly along the liberal--conservative spectrum---also
informs support for different types of AI moderation. Conservatives, who
tend to prioritize order, safety, and personal responsibility, may
support moderation that prevents harmful or illegal content. Liberals,
especially in recent years, may place greater emphasis on equity,
inclusivity, and social justice, and thus may be more supportive of
interventions that promote fairness (bias mitigation) or progressive
values (aspirational portrayals) (Chong et al., 2024; Chong \& Levy,
2018).

\begin{tcolorbox}[colback=gray!5!white, colframe=gray!80!black, title=Research Question 2]
How do individual-level factors---including personal experience
with AI, support for free speech, and political ideology---shape public
preferences for different goals of AI governance, such as accuracy and
reliability, safety, bias mitigation, and the promotion of aspirational
societal values?
\end{tcolorbox}

\subsubsection{Group-level involvement: Partisanship \&
Gender}\label{group-level-involvement-partisanship-gender}

People can also experience AI through the lens of group-level
involvement. Such involvement occurs when group membership shapes
exposure to AI technologies or to the societal debates surrounding their
regulation. This study examines two forms of group-level involvement:
political partisanship and gender.

\paragraph{\texorpdfstring{\emph{Partisanship}}{Partisanship}}\label{partisanship}

First, partisanship increasingly shapes attitudes toward digital
governance. In the United States, the issue of speech
moderation---particularly on digital platforms---has become highly
politicized. Republican political elites have framed moderation efforts
as ideologically biased and as threats to free speech (McCabe \& Kang,
2020). This discourse often targets progressive actors as overstepping
in regulating online content. As a result, we expect Republican
supporters to oppose forms of AI content moderation that are framed as
ideologically motivated---such as bias mitigation and the promotion of
aspirational portrayals. However, moderation in the name of safety may
find greater acceptance among Republicans, as it aligns with
conservative discourses of security and protection.

In contrast, Democratic partisans are more likely to view moderation as
a tool for fostering equity, inclusivity, and representation. We,
therefore, expect them to show higher support for AI moderation focused
on bias reduction and aspirational portrayals. In Germany, while digital
content governance is less politicized than in the U.S., the Green Party
has actively promoted strong regulatory measures to address
misinformation, hate speech, and inequality online (Hanfeld, 2025). As
such, we expect Green Party supporters to endorse AI content moderation
across all justifications. For other parties in Germany, where digital
policy debates are less polarized, we do not expect systematic
differences.

\paragraph{\texorpdfstring{\emph{Gender}}{Gender}}\label{gender}

Second, group-level involvement may arise from shared experiences of
harm or vulnerability. One such case is gender. Women are
disproportionately exposed to digital risks (De Ruiter, 2021; Wang \&
Kim, 2022). These experiences may sensitize them to the potential harms
of AI-generated content and increase their support for interventions
designed to moderate it. We, therefore, expect women to show greater
support for AI content moderation, particularly for justifications
centered on safety and bias reduction.

\begin{tcolorbox}[colback=gray!5!white, colframe=gray!80!black, title=Research Question 3]
How do group-level characteristics such as partisanship and gender
shape public preferences for different goals of AI governance, such as
accuracy and reliability, safety, bias mitigation, and the promotion of
aspirational societal values?
\end{tcolorbox}

\subsubsection{System-level involvement:
Country}\label{system-level-involvement-country}

Countries differ in their openness toward new technologies (Comin \&
Hobijn, 2010; Ding, 2024) and perceptions of technological risk (Douglas
\& Wildavsky, 1982). This is also evident when we examine the uses of
generative AI. In the United States---a country with a world-leading
digital technology sector and comparatively strong openness toward new
technologies---33 percent of respondents in a survey representative of
Americans adults said in August 2024 they had used AI-enabled chatbots
like ChaptGPT or Google Gemini (McClain et al., 2025). In contrast, in
September 2024 in Germany---a country without a strong digital
technology sector and more hesitant in its approach to new technology -
25 percent of respondents to a representative survey of Germans age 16
and older said they had used AI-enabled services like ChaptGPT or Google
Gemini (IfD-Allensbach, 2024). These figures illustrate that countries
differ in how citizens engage with emerging technologies.

Country-specific differences in AI use can also extend to attitudes
toward new technology and associated phenomena. For example, people vary
according to country in their views on the benefits and risks of AI
(Kelley et al., 2021). Similar differences can be found with regulatory
preferences for AI and digital technology (Rauchfleisch et al., 2025;
Riedl et al., 2021; Theocharis et al., 2025). We argue that these
national differences in engagement and preferences translate into
varying levels of societal involvement with AI, which we define as the
extent to which AI technologies are integrated into daily life,
institutional practices, and public debate.

We further assume that societal involvement conditions the role of
individual-level involvement. In highly involved societies, we expect
public opinion to be relatively uniform, such that highly and weakly
involved individuals hold similar views. In contrast, in less involved
societies, attitudes toward AI moderation should differ more strongly
depending on personal involvement. As such, in high-involvement
countries, low-involved individuals should stand out, while in
low-involvement countries, high-involved individuals should.

To examine these expectations, we compare public attitudes in the United
States (representing a high-involvement context) and Germany
(representing a low-involvement context).

\begin{tcolorbox}[colback=gray!5!white, colframe=gray!80!black, title=Research Question 4]
How does the system-level variable country shape public preferences
for different goals of AI governance, such as accuracy and reliability,
safety, bias mitigation, and the promotion of aspirational societal
values?
\end{tcolorbox}

\begin{tcolorbox}[colback=gray!5!white, colframe=gray!80!black, title=Research Question 5]
How does the relationship between individual- and group-level
involvement with AI and preferences for AI governance vary across
countries with differing levels of societal involvement with AI?
\end{tcolorbox}

\section{Methods}\label{methods}

We collected data in the U.S. and Germany over online panels. In the
U.S., 1,800 participants were recruited from the survey research company
Prolific (collected between 1 and 6 March 2024). We used a
representative quota sample for the US for sex, age, and political
affiliation (see Supplementary Information). Participants had to be
U.S.-based and aged 18 or older to participate in the study.
Participants were paid £0.75 (an hourly rate of £9; we ran the survey
through Prolific's European platform) for their study participation,
which took around 5 minutes to complete. 44 participants who failed a
simple attention check at the beginning of the study were excluded,
resulting in a sample of 1,756. On the starting page, we informed
participants about their rights (for example, that they could withdraw
from the study at any point by simply closing the browser) and asked
them to provide their consent. None of the questions asked for
personally identifiable information. In Germany, we also recruited 1,800
participants from the survey research company Bilendi (collected between
14 and 18 March 2024). We used a quota for age, gender, and regions in
Germany (16 states). The only difference from the US survey was that we
could directly filter out participants who failed the attention check
and thus ran the survey until we achieved 1,800 successful completes.

The descriptive statistics for all measured variables are reported in
Table 1 (for a complete table on the item level with the question
wording, see Supplementary Information). We measured AI moderation
preferences by asking respondents: ``Please indicate how important the
following criteria are for you when choosing AI-enabled services''. We
measured the evaluation on a 7-point scale (1 = not important at all; 7
= very important) for the four concepts related to AI moderation.
Accuracy and reliability (``The AI service consistently provides
accurate and reliable information or results based on its analysis and
data-driven insights.''), safety (``Measures are in place to prevent the
AI from generating or promoting illegal, dangerous,or harmful
content.''), bias mitigation (``Efforts are made to identify and reduce
biases in AI outputs, ensuring fairness and equity in treatment and
decision-making across different groups of people.''), and aspirational
imaginaries (``The AI aims to highlight and encourage positive societal
values, portraying an aspirational view of society.'').

\begin{longtable}{@{}
  >{\raggedright\arraybackslash}p{0.17\linewidth}
  >{\centering\arraybackslash}p{0.17\linewidth}
  >{\centering\arraybackslash}p{0.17\linewidth}
  >{\centering\arraybackslash}p{0.17\linewidth}
  >{\centering\arraybackslash}p{0.17\linewidth}@{}}
\caption{Descriptive statistics for all variables.}
\label{tab:descriptive} \\
\toprule
& \multicolumn{2}{c}{\textbf{US}} & \multicolumn{2}{c}{\textbf{Germany}} \\
\cmidrule(lr){2-3} \cmidrule(lr){4-5}
\textbf{Variable} & \textbf{M (SD)} & \textbf{n} & \textbf{M (SD)} & \textbf{n} \\
\midrule
\endfirsthead

\toprule
& \multicolumn{2}{c}{\textbf{US}} & \multicolumn{2}{c}{\textbf{Germany}} \\
\cmidrule(lr){2-3} \cmidrule(lr){4-5}
\textbf{Variable} & \textbf{M (SD)} & \textbf{n} & \textbf{M (SD)} & \textbf{n} \\
\midrule
\endhead

\bottomrule
\endfoot

Accuracy & 6.02 (1.28) & 1756 & 5.14 (1.66) & 1800 \\
Safety & 5.65 (1.68) & 1756 & 5.29 (1.75) & 1800 \\
Mitigating bias & 5.54 (1.61) & 1756 & 4.75 (1.71) & 1800 \\
Imaginaries & 4.55 (1.78) & 1756 & 4.43 (1.69) & 1800 \\
AI use ($\alpha$ US = 0.7; D = 0.80, Spearman-Brown US = 0.7; D = 0.80) & 3.18 (1.59) & 1756 & 2.70 (1.61) & 1800 \\
Free speech ($\alpha$ US = 0.9; D = 0.85, Spearman-Brown US = 0.9; D = 0.86) & 5.56 (1.34) & 1756 & 5.87 (1.16) & 1800 \\
Political orientation & 3.66 (1.86) & 1756 & 3.86 (1.15) & 1800 \\
Democratic Party/Green Party ID & 47.4\% & 1756 & 14.6\% & 1800 \\
Gender (female) & 50.3\% & 1756 & 50.0\% & 1800 \\
Education (high) & 17.4\% & 1756 & 19.6\% & 1800 \\
Age & 45.92 (15.94) & 1756 & 45.58 (15.48) & 1800 \\
\end{longtable}

AI use was measured with two items: one asking about AI use in the
professional or work context and one assessing AI use in personal life
and spare time (1 = never; 7 = very often). The two items were combined
into a mean index. Support for free speech was measured using two items
from Riedl et al.~(2021), which were adapted from Rojas et al.~(1996).
Political orientation was measured with a single scale (US=1-liberal;
7-conservative; Germany: 1-left; 7-right). To identify supporters of the
Democratic Party in the US and the Green Party in Germany, we recoded
the answers to a question about the general leaning toward a party in
the country.\footnote{The percentage of Democratic Party supporters is
  higher for this question than that reported for the party ID used for
  quota sampling, due to the wording of the question: ``In the US, many
  people lean towards a particular party for a long time, although they
  may occasionally vote for a different party. How about you, do you in
  general lean towards a particular party? If so, which one?'' This
  higher percentage is attributable to independents who lean towards the
  Democratic Party.} For education, we recoded responses into two
categories: ``postgraduate degree or higher'' and ``other'' (with
``other'' serving as the reference category).

As an analytical strategy, we use both datasets together for the
regression analysis. We first estimate, for each outcome, a model with
all predictors---including a country variable (Germany=0; US=1)---to
check whether single predictors have an overall association with the
outcome variable. We then estimate a second model in which we enter all
variables (mean-centered) as interaction terms with the country
variable. This also allows us to test, through the interactions, whether
there is a country difference in the explanatory strength of the
predictors. A positive estimate for the interaction term would indicate
that the predictor is stronger in the US, whereas a negative estimate
would indicate a stronger predictor in Germany. As these interaction
terms are difficult to interpret, we will visualize them as marginal
effect plots in the results section (we also report single-country
regression models in Supplementary Information).

\section{Results}\label{results}

\subsection{RQ1 Preferences}\label{rq1-preferences}

We start with a descriptive analysis of respondents' functional
preferences when choosing AI-enabled services. Figure 1 displays the
distribution of responses in Germany and the United States when
participants were asked how important each of these criteria is in
selecting an AI-enabled service.

\begin{figure}[htbp]
  \centering
  \includegraphics[width=\linewidth]{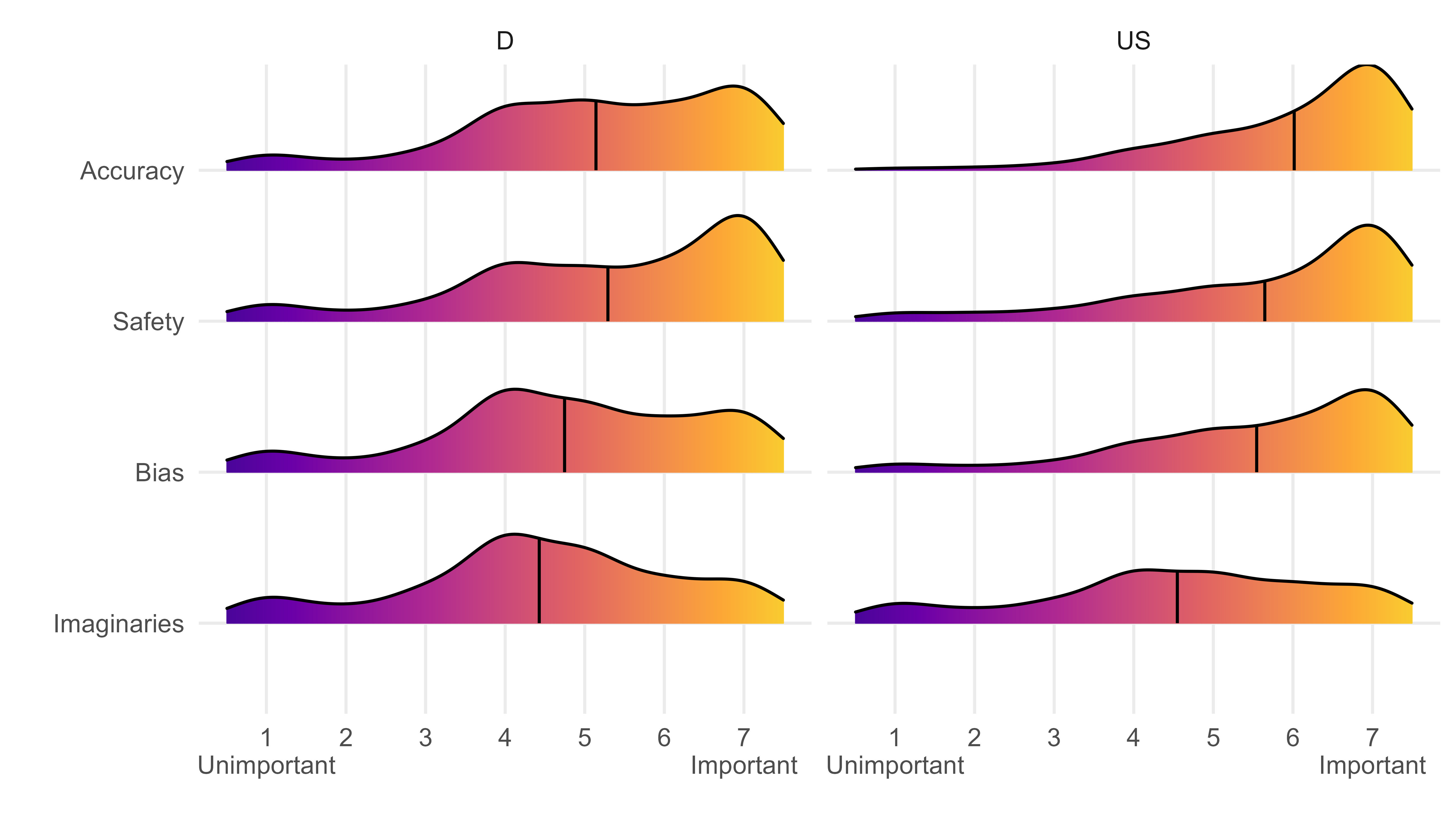}
  \caption{Distribution for the four outcome variables for Germany (left) and the US (right). Vertical lines indicate the mean.}
  \label{fig:distribution1}
\end{figure}

As expected, support varies systematically across approaches. Public
support is strongest for moderation goals oriented toward accuracy,
reliability, and safety. Both U.S. and German respondents assign high
importance to accuracy. Similarly, preventing harm---defined as
preventing the generation of illegal, dangerous, or harmful content---is
widely supported. These safety-oriented adjustments to model outputs
appear to be largely noncontroversial, likely due to their alignment
with conventional risk regulation in digital communication environments.

Support declines, however, as moderation shifts from safety to more
normative goals. Bias mitigation, which aims to promote fairness and
equity, is still positively received but exhibits more variation,
especially among German respondents.

Importantly, country-level differences in support patterns (see also the
results for country in Figure 2) align with broader national trends in
technology adoption and risk perception. A Welch Two-Sample t-test
indicates a significant difference (t(3553.4) = --8.98, p \textless{}
.001) in AI use between the US and Germany. Participants in the
U.S.--home to a world-leading digital technology sector and generally
higher openness to new technologies (Comin \& Hobijn, 2010; Ding,
2024)--reported higher AI use scores (M = 3.18, SD = 1.59) than those in
Germany (M = 2.70, SD = 1.61)--where adoption of generative AI tools
remains more cautious and public discourse often emphasizes potential
risks. Furthermore, 26.8\% of respondents in Germany reported that they
never use AI-supported applications for either personal or professional
purposes, compared to only 10.8\% in the US.

These differences in usage correspond with differences in preference.
The United States consistently shows greater support across all
categories. In contrast, Germany shows more reserved or varied support,
particularly for normatively driven moderation goals. These differences
reflect varying levels of societal involvement with AI,

\subsection{Explaining preferences for
AI-moderation}\label{explaining-preferences-for-ai-moderation}

We now examine how different explanatory factors shape preferences for
the functional features of AI-enabled systems among respondents in
Germany and the United States. Figure 2 displays the estimated
coefficients from our model (for the complete tables for the models, see
Supplementary information).

\begin{figure}[htbp]
  \centering
  \includegraphics[width=\linewidth]{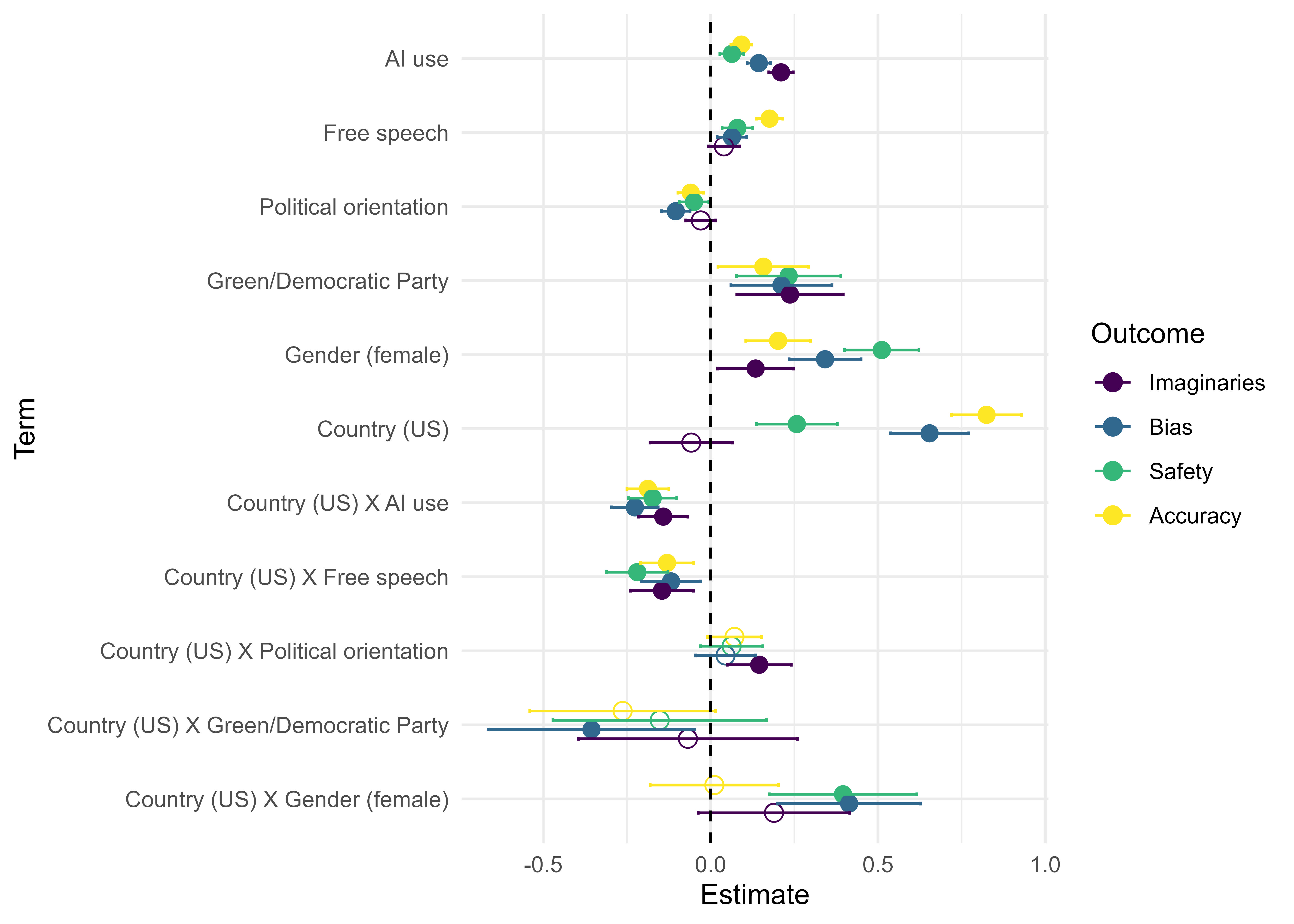}
  \caption{Estimates with 95\%-CIs for all four outcome variables.
Non-significant estimates are indicated as empty dots. Significant
single estimates indicate an overall association. Significant negative
interaction estimates indicate that the association is relatively
stronger in Germany. Significant positive interaction estimates the
association is relatively stronger in the U.S.}
  \label{fig:distribution2}
\end{figure}

\subsubsection{RQ2 Individual-Level
Factors}\label{rq2-individual-level-factors}

At the individual level, personal experience with AI consistently
predicts support for all tested features across both countries. The
association is particularly strong for moderation goals that are
specific to AI systems, such as bias mitigation ($\beta$ = 0.14, p \textless{}
.001 0, 95\% CI {[}0.11, 0.18{]}) and aspirational portrayals ($\beta$ = 0.21,
p \textless{} .001 0, 95\% CI {[}0.17, 0.25{]}). This indicates that
direct experience with AI fosters a more nuanced understanding of its
societal implications, making individuals more receptive to content
interventions targeting AI-specific harms and potentials.

We also find a significant role for free speech attitudes, though the
pattern is somewhat counterintuitive. Overall, stronger support for free
speech is associated with greater support for accuracy ($\beta$ = 0.18, p
\textless{} .001, 95\% CI {[}0.14, 0.22{]}), safety-related moderation
($\beta$ = 0.08, p \textless{} .001, 95\% CI {[}0.03, 0.13{]}), and bias
reduction ($\beta$ = 0.06, p = .005, 95\% CI {[}0.02, 0.11{]}), but not for
the promotion of aspirational imaginaries ($\beta$ = 0.04, p = .095, 95\% CI
{[}-0.01, 0.09{]}). Political ideology also aligns with our
expectations: in both countries, individuals who identify as liberal or
left are more supportive of AI moderation than those who identify as
conservative. Only for aspirational portrayal, the estimate is not
significant ($\beta$ = -0.03, p = .201, 95\% CI {[}-0.07, 0.02{]}).

\subsubsection{RQ3 Group-Level
Involvement}\label{rq3-group-level-involvement}

Turning to group-level factors, we observe that partisan alignment plays
a role consistent with party cues. Overall, self-identified supporters
of the Green Party (Germany) and the Democratic Party (U.S.) are more
likely to support all four forms of AI moderation than supporters of
other parties (see Figure 3). These patterns mirror the elite discourse
within these parties, suggesting that elite signaling helps structure
public attitudes toward AI governance.

We also find that gender shapes moderation preferences, as women see all
four forms of AI moderation as important. Primarily for safety ($\beta$ =
0.51, p \textless{} .001, 95\% CI {[}0.40, 0.62{]}) and bias reduction
($\beta$ = 0.34, p \textless{} .001, 95\% CI {[}0.23, 0.45{]}), women, on
average, are more likely than men to support interventions. This
supports the argument that group-level exposure to digital risks---such
as online harassment and misrepresentation---can translate into greater
support for protective content interventions.

\subsubsection{RQ4 Cross-National
Differences}\label{rq4-cross-national-differences}

The most pronounced differences between countries emerge for preferences
related to accuracy and reliability ($\beta$ = 0.82, p \textless{} .001, 95\%
CI {[}0.72, 0.93{]}), safety ($\beta$ = 0.26, p \textless{} .001, 95\% CI
{[}0.14, 0.38{]}), and bias mitigation ($\beta$ = 0.65, p \textless{} .001,
95\% CI {[}0.54, 0.77{]}). As expected, respondents in the U.S.---a
context characterized by higher societal involvement with AI---express
stronger support for these types of moderation compared to respondents
in Germany. This finding supports the idea that greater societal
exposure to AI corresponds with increased public demand for moderation
practices tailored to the specific risks and opportunities associated
with these systems.

Interestingly, this pattern does not extend to support for aspirational
portrayals ($\beta$ = -0.06, p = .358, 95\% CI {[}-0.18, 0.07{]})---that is,
AI-generated content promoting particular visions of society. For this
type of moderation, we observe no significant difference between the two
countries. This suggests that support for aspirational content
moderation may be driven more by ideological values than by levels of
societal AI involvement.

\subsubsection{RQ5 Involvement Differences by
Country}\label{rq5-involvement-differences-by-country}

Figure 2 also illustrates interaction terms from our second model, which
indicate whether the influence of a given variable differs significantly
across countries. Negative interaction estimates suggest stronger
association in Germany; positive estimates suggest stronger association
in the U.S.

We find that individual-level differences in AI use have a significant
impact in Germany but a considerably smaller effect in the U.S. The same
is true for support for free speech: these attitudes are more predictive
of preferences in Germany than in the U.S. In contrast, political
ideology only has a differential association in the U.S., where it
significantly predicts support for aspirational portrayals ($\beta$ = 0.15, p
= .003, 95\% CI {[}0.05, 0.24{]}). This suggests that in the U.S., the
political discourse around aspirational imaginaries of
society---particularly those embedded in AI systems---is more developed
and polarized.

Regarding group-level variables, partisan differences do not exhibit
consistent cross-national interaction terms with only a difference
between countries for mitigating bias ($\beta$ = -0.36, p = .024, 95\% CI
{[}-0.66, -0.05{]}) as the association is stronger in Germany. For
gender, however, the associations are more pronounced in the US for
preferences related to safety ($\beta$ = 0.40, p \textless{} .001, 95\% CI
{[}0.18, 0.62{]}) and bias mitigation ($\beta$ = 0.41, p \textless{} .001,
95\% CI {[}0.20, 0.63{]}).

\begin{figure}[htbp]
  \centering
  \includegraphics[width=\linewidth]{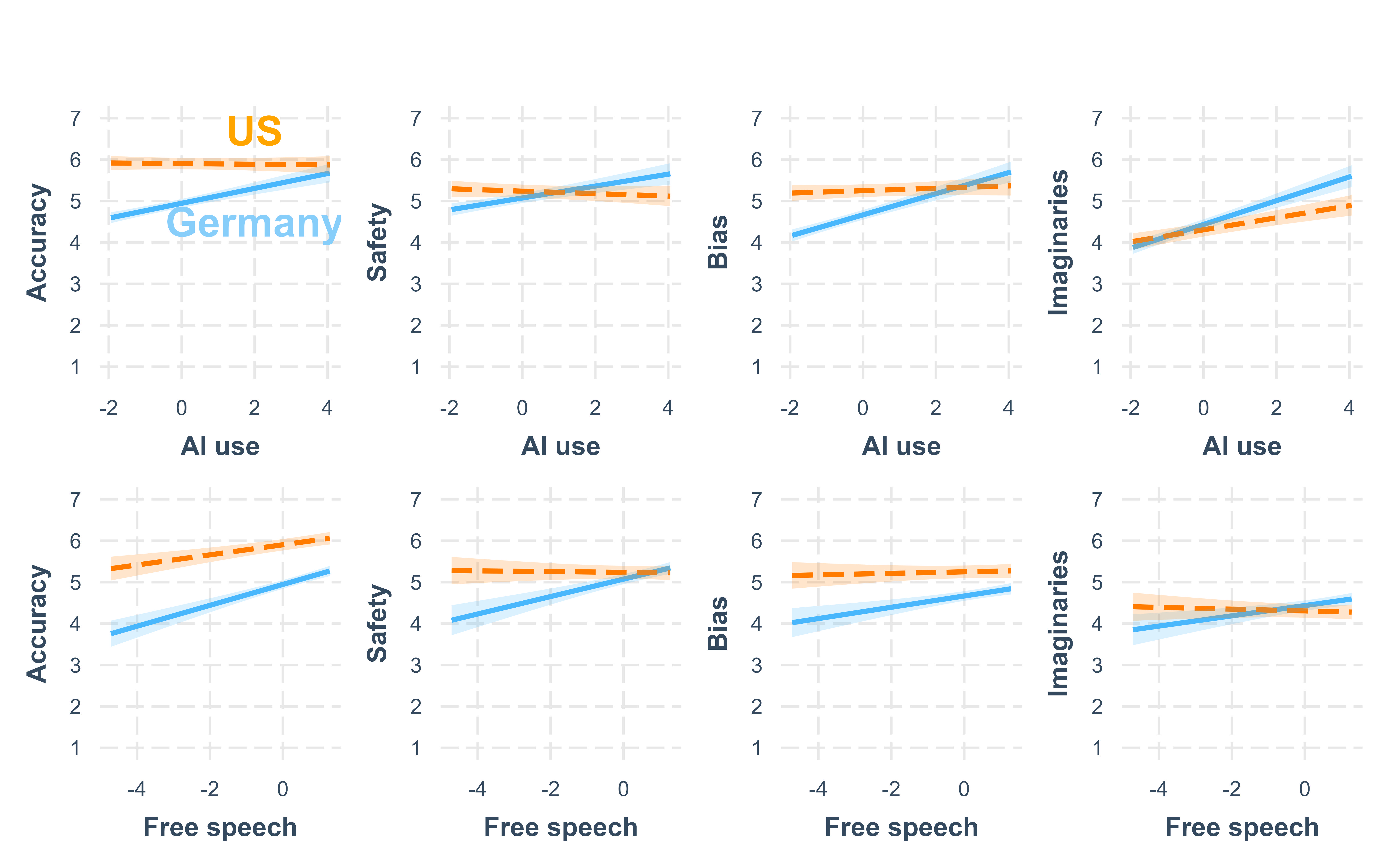}
  \caption{Interactions for AI use and free speech with country for all
four outcome variables. All interaction terms are significant.}
  \label{fig:distribution3}
\end{figure}

Figure 3 further illustrates these dynamics by visualizing the
interactions of the country variable with AI use and free speech
attitudes. The figure shows that in the U.S., there is little difference
between individuals with high and low AI experience in terms of their
moderation preferences. In Germany, however, these differences are much
more pronounced---particularly among respondents with low AI experience,
who are significantly less supportive of moderation. As AI experience
increases, the gap between German and U.S. respondents narrows.

A similar pattern emerges for free speech attitudes. Again, this
supports our broader argument: in high-involvement societies like the
U.S., attitudes toward AI moderation are more uniformly developed,
reducing the explanatory power of individual-level variation. In
contrast, in low-involvement societies, such as Germany, individual
experiences and values play a larger role in shaping
attitudes.\footnote{We also report single models for each country and
  outcome variable in the Supplementary Materials. They support the
  overall interpretations of the analysis with the interaction terms. There, we also report a specification curve analysis }

\section{Discussion}\label{discussion}

This study brings a comparative and attitudinal perspective to the
debate on AI alignment by examining how users evaluate key features of
AI-enabled systems. We show that individuals hold distinct preferences
for moderation mechanisms that influence model outputs and that these
preferences vary systematically between countries. Respondents in the
United States report significantly higher levels of AI use than those in
Germany, reflecting broader national differences in technology adoption
and societal engagement with AI. Across both contexts, accuracy,
reliability, and safety receive the strongest public
support---indicating a shared baseline of expectations for trustworthy
and safe systems. In contrast, support is more conditional for
interventions to mitigate bias or promote aspirational societal values,
especially among German respondents. This aligns with the view that bias
correction involves normative judgments that can be politically charged
and subject to disagreement. Similarly, the lower support of
interventions promoting aspirational imaginaries indicates hesitancy
about the active role of AI in shaping cultural narratives---and
potentially associated concerns about legitimacy, ideological overreach,
and value alignment. U.S. participants consistently express higher
support across all dimensions, which aligns with their greater exposure
to and involvement with AI technologies. The difference in support
between Germany and the U.S. could be an expression of the maturity of
the public discourse and awareness about the functioning of AI-enabled
systems, indicating a lower awareness among German respondents about
related issues.

Our analysis further reveals that personal experience with AI, political
ideology, gender, and support for free speech shape attitudes toward AI
alignment---but with varying strength across countries. In Germany, both
AI experience and free speech attitudes are stronger predictors of
support, suggesting that in contexts with lower exposure,
individual-level factors play a more decisive role. In the United
States, where AI technologies are more deeply embedded in public and
institutional life, views on AI moderation appear more consolidated,
with political ideology particularly influencing support for
aspirational interventions.

Our finding on the role of free speech support on preferences for
adjustments of model outputs is especially interesting since it
contrasts previous findings on digital content moderation, where free
speech concerns often predict more resistance to moderation
interventions by companies or states (Jang et al., 2024; Rauchfleisch \&
Jungherr, 2024). One possible interpretation is that respondents do not
view generative AI output as equivalent to human speech. That is, the
normative privilege of free speech may not extend, in the public's view,
to AI-generated content. These findings suggest that assumptions from
earlier debates about digital content moderation cannot be automatically
transferred to the case of AI. Future policy and public debate on AI
moderation should take these differences into account.

Our study is subject to several important limitations. First, there is a
temporal dimension to consider. As AI-enabled systems become more
prevalent in daily life, both individual experience with these
technologies and public discourse around them are likely to evolve. The
cross-national differences we identify may, therefore, be time-bound and
could diminish over time as country-level involvement with AI converges
internationally.

Second, our analysis is limited to just two countries. Future research
should broaden the comparative scope to include a more diverse set of
countries, particularly those with varying levels of technological
integration and public attitudes toward AI. In this context, we see
particular value in examining countries in Asia, where both the pace and
form of AI adoption differ substantially from Western contexts.
Moreover, our operationalization of ``technology involvement'' is
relatively coarse. Future studies should develop and test more nuanced
and systematic measures---such as indicators of public discourse,
regulatory activity, or the economic significance of AI in a given
country.

Finally, our research design is cross-sectional and based on
self-reported data. This limits the causal inferences that can be drawn
and may be subject to bias in participants' self-assessments of AI use
and preferences. Future work should incorporate more objective measures
of AI experience and leverage experimental or longitudinal designs to
capture how individuals respond to concrete AI interventions rather than
relying solely on abstract descriptions or stated preferences.

Our findings point to a set of important and more general considerations
that should be taken into account and pursued further. This includes
consideration of geopolitical competition and conflict, the role of
companies, and the deep opaqueness and unassessability in the pipeline
in model provision.

Our findings highlight substantial cross-national differences in public
attitudes toward AI. These differences are particularly significant in
today's AI landscape, where U.S. or Chinese companies develop the most
widely used systems. As a result, public expectations for AI moderation
may not only reflect concerns about functionality and fairness, but also
the perceived degree of foreign versus domestic control over digital
environments. Similar to dynamics observed in international trade
(Jungherr et al., 2018), attitudes toward the countries of origin of AI
technologies can ``contaminate'' perceptions of the technologies
themselves. This dimension warrants close attention in a geopolitical
climate marked by intense competition and strategic rivalry.

Moreover, access to AI models is shaped by the strategic and commercial
interests of the companies that develop them. Whether driven by profit
or geopolitical considerations, these motivations may influence both the
design and availability of AI systems in ways that affect public trust.
Importantly, AI moderation is just one step in a longer, largely opaque
chain of decisions made during model development, training, deployment,
and evaluation. At each stage, political values---intentionally or
not---may become embedded in technical systems. To ensure global
legitimacy and public trust in AI, these decision-making chains must
become more transparent, assessable, and, where appropriate, open to
public negotiation and contestation.

Currently, model training and moderation practices remain largely hidden
from public scrutiny. This lack of visibility risks eroding public
confidence and enabling politicized narratives about AI bias or hidden
agendas. In a context of growing diversity in AI development---spanning
open-source and commercial models, varying origins, and geopolitical
alignments (Buyl et al., 2024)---there is an urgent need for a more
mature, structured debate about legitimate approaches to model
adjustment, both during training and in real-time operation.

Without such a debate, we risk stumbling from one controversy to the
next, fostering a general climate of suspicion toward AI. Transparency
alone is not enough; societies must also articulate clear expectations
of what they want from AI systems and how those systems should be
governed. Therefore, companies, policymakers, and researchers must take
active responsibility for documenting and debating the principles,
procedures, and techniques underpinning justified AI moderation. As
O'Neill (2021) argues, digital systems must be made assessable to users.
If AI moderation practices remain opaque, public trust will
deteriorate---especially when high-profile errors are framed as evidence
of hidden political or cultural agendas.

Ultimately, realizing the societal benefits of AI will depend on
building a public governance framework that allows for visibility,
legitimacy, and accountability in model development and moderation.
Failing to do so risks deepening skepticism and undermining AI's
long-term viability in democratic societies.

\section{Funding sources}\label{funding-sources}

A. Rauchfleisch's work was supported by the National Science and
Technology Council, Taiwan (R.O.C) (Grant No 113-2628-H-002-018-).

\section{Declaration of interests}\label{declaration-of-interests}

A. Jungherr and A. Rauchfleisch declare no potential competing interests
with respect to the research, authorship, and/or publication of this
article.

\section{Author Information}\label{author-information}

\textbf{Andreas Jungherr} holds the Chair for Political Science, especially Digital Transformation, at the University of Bamberg. He examines the impact of digital media on politics and society, with a special focus on algorithms, artificial intelligence, and governance. He is the author of \textit{Retooling Politics: How Digital Media is Shaping Democracy} (with Gonzalo Rivero and Daniel Gayo-Avello, Cambridge University Press, 2020) and \textit{Digital Transformations of the Public Arena} (with Ralph Schroeder, Cambridge University Press, 2022).

\medskip

\textbf{Adrian Rauchfleisch} is an Associate Professor at the Graduate Institute of Journalism, National Taiwan University. His research focuses on the interplay of politics, technology, and journalism in Asia, Europe, and the United States. His new project explores Artificial Intelligence's influence on society across different cultural contexts.



\newpage{}

\section{Appendix: Supplementary Materials}\label{sec-app_appendix}

\appendix

\subsection{Data}

\subsubsection{Sample and population data US}
\label{section:pop}
We recruited participants via the platform Prolific and used their predefined quotas for the U.S. population regarding age, gender, and party identification (Republican, Democrat, or Independent). In terms of gender, our sample consists of 48.75\% female, 49.66\% male, and 1.59\% identifying as other. Regarding party affiliation, 31.44\% are Democrats, 42.71\% are Independents, and 25.85\% are Republicans. We also achieved a good distribution across the different age brackets (see Table \ref{table:age_bracket_distribution}). The percentage of Democratic Party supporters is higher for the variable used in our model than that reported for the party ID used for the quota sampling, due to the wording of the question: “In the US, many people lean toward a particular party for a long time, although they may occasionally vote for a different party.  How about you, do you in general lean towards a particular party? If so, which one?” This higher percentage is attributable to independents who lean towards the Democratic Party. 

\begin{table}[H]
\centering
\begin{tabular}[t]{lrr}
\toprule
\textbf{Age Bracket} & \textbf{Count} & \textbf{Percentage (\%)} \\
\midrule
18-27     & 310  & 17.65 \\
28-37     & 314  & 17.88 \\
38-47     & 291  & 16.57 \\
48-57     & 296  & 16.86 \\
58-94    & 545  & 31.04 \\
\bottomrule
\end{tabular}
\caption{Distribution of Sample Across Age Brackets.}
\label{table:age_bracket_distribution}
\end{table}

\subsubsection{Sample and population data Germany}
For Germany, we used quota calculated by the survey research company Bilendi based on Eurostat statistics for Germany. Our sample matched the distribution for Gender, age brackets, and regions (16 states). For an overview see Table \ref{table:germany_desc}

\begin{table}[H]
\centering
\begin{tabular}[t]{lrr}
\toprule
\textbf{Category} & \textbf{Count} & \textbf{Percentage (\%)} \\
\midrule
18-29     & 361  & 20 \\
30-39     & 336  & 19 \\
40-49     & 324  & 18 \\
50-59     & 433  & 24 \\
60+    & 343  & 19 \\
Female & 900 & 50 \\
Male & 900 & 50 \\
Baden-Württemberg  & 232            & 13                       \\
Bayern             & 286            & 16                       \\
Berlin             & 85             & 5                        \\
Brandenburg        & 53             & 3                        \\
Bremen             & 18             & 1                        \\
Hamburg            & 37             & 2                        \\
Hessen             & 140            & 8                        \\
Mecklenburg-Vorpommern & 35         & 2                        \\
Niedersachsen      & 179            & 10                       \\
Nordrhein-Westfalen& 395            & 22                       \\
Rheinland-Pfalz    & 89             & 5                        \\
Saarland           & 18             & 1                        \\
Sachsen            & 89             & 5                        \\
Sachsen-Anhalt     & 53             & 3                        \\
Schleswig-Holstein & 53             & 3                        \\
Thüringen          & 38             & 2                        \\

\bottomrule
\end{tabular}
\caption{Distribution of Sample Across Age Brackets, gender, and region.}
\label{table:germany_desc}
\end{table}

\subsection{Measures}

\begin{table}[H]
\begin{tabular}{L{4cm}L{4cm}cccc}
  \toprule
  \textbf{Variable} 
    & \textbf{Question Wording / Operationalization} 
    & \multicolumn{2}{c}{\textbf{US}}
    & \multicolumn{2}{c}{\textbf{Germany}} \\
  \cmidrule(lr){3-4} \cmidrule(lr){5-6}
      &  & M (SD) & n  & M (SD) & n \\
  \midrule
Importance accuracy & The AI service consistently provides accurate and reliable information or results based on its analysis and data-driven insights. (1="Not important at all", 7="Very important") & 6.02 (1.28) & 1756 & 5.14 (1.66) & 1800\\

Importance safety & Measures are in place to prevent the AI from generating or promoting illegal, dangerous, or harmful content.. (1="Not important at all", 7="Very important") & 5.65 (1.68) & 1756 & 5.29 (1.75) & 1800\\

Importance bias mitigation & Efforts are made to identify and reduce biases in AI outputs, ensuring fairness and equity in treatment and decision-making across different groups of people. (1="Not important at all", 7="Very important") & 5.54 (1.61) & 1756 & 4.75 (1.71) & 1800\\

Importance aspirational imaginaries & The AI aims to highlight and encourage positive societal values, portraying an aspirational view of society. (1="Not important at all", 7="Very important") & 4.55 (1.78) & 1756 & 4.43 (1.69) & 1800\\

AI use (2 items, $\alpha$ US = 0.7; D = 0.80, Spearman-Brown US = 0.7; D = 0.80) & (1="never", 7="very often") & 3.18 (1.59) & 1756 & 2.7 (1.61) & 1800\\
 & How frequently do you use AI-supported applications or services in your...professional or work environment & 2.96 (1.93) & 1756 & 2.59 (1.80) & 1800\\
 & ...personal life and spare time & 3.41 (1.70) & 1756 & 2.82 (1.72) & 1800\\

\bottomrule
\end{tabular}%
\caption{First part with descriptive statistics for all relevant variables by country (US and Germany).}
\label{tab:descr1}
\end{table}

\begin{table}[H]
\begin{tabular}{L{4cm}L{4cm}cccc}
  \toprule
  \textbf{Variable} 
    & \textbf{Question Wording / Operationalization} 
    & \multicolumn{2}{c}{\textbf{US}}
    & \multicolumn{2}{c}{\textbf{Germany}} \\
  \cmidrule(lr){3-4} \cmidrule(lr){5-6}
      &  & M (SD) & n  & M (SD) & n \\
  \midrule

Political orientation & (1-"liberal", 7="conservative") & 3.66 (1.86) & 1756 & 3.86 (1.15) & 1800\\
Support for free speech (2 items, $\alpha$ US = 0.9; D = 0.85, Spearman-Brown US = 0.9; D = 0.86) & (1="Strongly disagree", 7="Strongly agree") & 5.56 (1.34) & 1756 & 5.87 (1.16) & 1800\\
 & Everybody should have the freedom to publicly say what they believe to be true. & 5.69 (1.35) & 1756 & 6.01 (1.18) & 1800\\
 & No matter how controversial an idea is, an individual should be able to express it publicly. & 5.44 (1.46) & 1756 & 5.72 (1.30) & 1800\\

Democratic Party/Green Party ID &  & 47.4\% & 1756 & 14.6\% & 1800\\

Gender Female & 1=female/other;0=male & 50.3\% & 1756 & 50.0\% & 1800\\

Education (high) & 1=Postgraduate or professional degree, including master's, doctorate, medical or law degree (e.g., MA, MS, PhD, JD, graduate school);0=other & 17.4\% & 1756 & 19.6\% & 1800\\

\bottomrule
\end{tabular}%
\caption{Second part with descriptive statistics for all relevant variables by country (US and Germany).}
\label{tab:descr2}
\end{table}

\subsection{Model results}
In the first sub-section, we report the models that we report in the main paper. In the second subsection, we report the single country models.
\subsubsection{Tables for models reported in the paper}
In this section, we show the complete models reported in the main paper. For each outcome variable, we estimate one model. The estimates reported for the single variables are from the models without interaction effects. The interaction effects are from the models with interaction effects.

\paragraph{Single variable models}
\begin{table}[H]
\begin{tabular}[t]{lrrrrl}
\toprule
Predictors & Estimate & SE & LL & UL & p\\
\midrule
Intercept & 4.96 &  & 4.87 & 5.05 & <0.001\\
Political orientation & -0.06 &  & -0.10 & -0.02 & 0.002\\
Country (US) & 0.82 &  & 0.72 & 0.93 & <0.001\\
Free speech & 0.18 &  & 0.14 & 0.22 & <0.001\\
Gender (female) & 0.20 &  & 0.10 & 0.30 & <0.001\\
Education (high) & 0.30 &  & 0.18 & 0.43 & <0.001\\
AI use & 0.09 &  & 0.06 & 0.12 & <0.001\\
Age & 0.01 &  & 0.00 & 0.01 & <0.001\\
Green/Democratic Party & 0.16 &  & 0.02 & 0.29 & 0.023\\
\midrule
Observations & 3556 &  &  &  & \\
R2 / R2 adjusted & 0.126 / 0.124 &  &  &  & \\
\bottomrule
\end{tabular}
\caption{Linear regression model results with 95\% CIs. The outcome variable is importance of accuracy.}
\end{table}

\begin{table}[H]
\begin{tabular}[t]{lrrrrl}
\toprule
Predictors & Estimate & SE & LL & UL & p\\
\midrule
Intercept & 4.99 &  & 4.89 & 5.09 & <0.001\\
Political orientation & -0.05 &  & -0.09 & -0.01 & 0.028\\
Country (US) & 0.26 &  & 0.14 & 0.38 & <0.001\\
Free speech & 0.08 &  & 0.03 & 0.13 & 0.001\\
Gender (female) & 0.51 &  & 0.40 & 0.62 & <0.001\\
Education (high) & 0.11 &  & -0.03 & 0.25 & 0.139\\
AI use & 0.06 &  & 0.03 & 0.10 & 0.001\\
Age & 0.02 &  & 0.01 & 0.02 & <0.001\\
Green/Democratic Party & 0.23 &  & 0.08 & 0.39 & 0.003\\
\midrule
Observations & 3556 &  &  &  & \\
R2 / R2 adjusted & 0.066 / 0.064 &  &  &  & \\
\bottomrule
\end{tabular}
\caption{Linear regression model results with 95\% CIs. The outcome variable is importance of safety.}
\end{table}

\begin{table}[H]
\begin{tabular}[t]{lrrrrl}
\toprule
Predictors & Estimate & SE & LL & UL & p\\
\midrule
Intercept & 4.57 &  & 4.47 & 4.67 & <0.001\\
Political orientation & -0.10 &  & -0.15 & -0.06 & <0.001\\
Country (US) & 0.65 &  & 0.54 & 0.77 & <0.001\\
Free speech & 0.06 &  & 0.02 & 0.11 & 0.005\\
Gender (female) & 0.34 &  & 0.23 & 0.45 & <0.001\\
Education (high) & 0.08 &  & -0.06 & 0.22 & 0.265\\
AI use & 0.14 &  & 0.11 & 0.18 & <0.001\\
Age & 0.01 &  & 0.01 & 0.02 & <0.001\\
Green/Democratic Party & 0.21 &  & 0.06 & 0.36 & 0.006\\
\midrule
Observations & 3556 &  &  &  & \\
R2 / R2 adjusted & 0.105 / 0.103 &  &  &  & \\
\bottomrule
\end{tabular}
\caption{Linear regression model results with 95\% CIs. The outcome variable is importance of bias mitigation.}
\end{table}

\begin{table}[H]
\begin{tabular}[t]{lrrrrl}
\toprule
Predictors & Estimate & SE & LL & UL & p\\
\midrule
Intercept & 4.38 &  & 4.28 & 4.49 & <0.001\\
Political orientation & -0.03 &  & -0.07 & 0.02 & 0.201\\
Country (US) & -0.06 &  & -0.18 & 0.07 & 0.358\\
Free speech & 0.04 &  & -0.01 & 0.09 & 0.095\\
Gender (female) & 0.13 &  & 0.02 & 0.25 & 0.020\\
Education (high) & -0.04 &  & -0.19 & 0.10 & 0.553\\
AI use & 0.21 &  & 0.17 & 0.25 & <0.001\\
Age & 0.01 &  & 0.00 & 0.01 & <0.001\\
Green/Democratic Party & 0.24 &  & 0.08 & 0.40 & 0.003\\
\midrule
Observations & 3556 &  &  &  & \\
R2 / R2 adjusted & 0.042 / 0.040 &  &  &  & \\
\bottomrule
\end{tabular}
\caption{Linear regression model results with 95\% CIs. The outcome variable is importance of showing aspirational version of the world.}
\end{table}

\paragraph{Models with interaction terms}
\begin{table}[H]
\begin{tabular}[t]{lrrrrl}
\toprule
Predictors & Estimate & SE & LL & UL & p\\
\midrule
Intercept & 4.94 &  & 4.84 & 5.05 & <0.001\\
Political orientation & -0.12 &  & -0.19 & -0.06 & <0.001\\
Country (US) & 0.96 &  & 0.78 & 1.13 & <0.001\\
Free speech  & 0.25 &  & 0.19 & 0.31 & <0.001\\
Gender (female) & 0.17 &  & 0.04 & 0.31 & 0.013\\
Education (high) & 0.43 &  & 0.26 & 0.60 & <0.001\\
AI use & 0.18 &  & 0.13 & 0.23 & <0.001\\
Age & 0.00 &  & -0.00 & 0.01 & 0.311\\
Green/Democratic Party & 0.30 &  & 0.11 & 0.50 & 0.002\\
Political orientation X Country (US) & 0.07 &  & -0.01 & 0.15 & 0.085\\
Country (US) X Free speech & -0.13 &  & -0.21 & -0.05 & 0.001\\
Country (US) X Gender (female)  & 0.01 &  & -0.18 & 0.20 & 0.909\\
Country (US) X Education (high) & -0.34 &  & -0.59 & -0.09 & 0.007\\
Country (US) X AI use & -0.19 &  & -0.25 & -0.12 & <0.001\\
Country (US) X Age & 0.01 &  & 0.01 & 0.02 & 0.001\\
Country (US) X Green/Democratic Party & -0.26 &  & -0.54 & 0.01 & 0.063\\
\midrule
Observations & 3556 &  &  &  & \\
R2 / R2 adjusted & 0.146 / 0.142 &  &  &  & \\
\bottomrule
\end{tabular}
\caption{Linear regression model results with 95\% CIs. The outcome variable is importance of accuracy.}
\end{table}

\begin{table}[H]
\begin{tabular}[t]{lrrrrl}
\toprule
Predictors & Estimate & SE & LL & UL & p\\
\midrule
Intercept & 5.07 &  & 4.95 & 5.19 & <0.001\\
Political orientation & -0.10 &  & -0.17 & -0.03 & 0.004\\
Country (US) & 0.16 &  & -0.04 & 0.36 & 0.107\\
Free speech  & 0.21 &  & 0.14 & 0.28 & <0.001\\
Gender (female) & 0.29 &  & 0.14 & 0.45 & <0.001\\
Education (high) & 0.21 &  & 0.02 & 0.41 & 0.034\\
AI use & 0.14 &  & 0.09 & 0.20 & <0.001\\
Age & 0.01 &  & 0.00 & 0.02 & <0.001\\
Green/Democratic Party & 0.31 &  & 0.09 & 0.54 & 0.006\\
Political orientation X Country (US) & 0.06 &  & -0.03 & 0.16 & 0.187\\
Country (US) X Free speech & -0.22 &  & -0.31 & -0.13 & <0.001\\
Country (US) X Gender (female)  & 0.40 &  & 0.18 & 0.62 & <0.001\\
Country (US) X Education (high) & -0.30 &  & -0.58 & -0.01 & 0.040\\
Country (US) X AI use & -0.17 &  & -0.25 & -0.10 & <0.001\\
Country (US) X Age & 0.01 &  & 0.01 & 0.02 & <0.001\\
Country (US) X Green/Democratic Party & -0.15 &  & -0.47 & 0.17 & 0.350\\
\midrule
Observations & 3556 &  &  &  & \\
R2 / R2 adjusted & 0.088 / 0.085 &  &  &  & \\
\bottomrule
\end{tabular}
\caption{Linear regression model results with 95\% CIs. The outcome variable is importance of safety.}
\end{table}

\begin{table}[H]
\begin{tabular}[t]{lrrrrl}
\toprule
Predictors & Estimate & SE & LL & UL & p\\
\midrule
Intercept & 4.67 &  & 4.55 & 4.78 & <0.001\\
Political orientation & -0.16 &  & -0.23 & -0.09 & <0.001\\
Country (US) & 0.58 &  & 0.39 & 0.78 & <0.001\\
Free speech  & 0.14 &  & 0.07 & 0.20 & <0.001\\
Gender (female) & 0.11 &  & -0.04 & 0.26 & 0.136\\
Education (high) & 0.12 &  & -0.07 & 0.31 & 0.231\\
AI use & 0.26 &  & 0.20 & 0.31 & <0.001\\
Age & 0.01 &  & 0.00 & 0.02 & <0.001\\
Green/Democratic Party & 0.40 &  & 0.18 & 0.62 & <0.001\\
Political orientation X Country (US) & 0.04 &  & -0.05 & 0.13 & 0.332\\
Country (US) X Free speech & -0.12 &  & -0.21 & -0.03 & 0.009\\
Country (US) X Gender (female)  & 0.41 &  & 0.20 & 0.63 & <0.001\\
Country (US) X Education (high) & -0.17 &  & -0.45 & 0.10 & 0.214\\
Country (US) X AI use & -0.23 &  & -0.30 & -0.16 & <0.001\\
Country (US) X Age & 0.01 &  & 0.00 & 0.02 & 0.020\\
Country (US) X Green/Democratic Party & -0.36 &  & -0.66 & -0.05 & 0.024\\
\midrule
Observations & 3556 &  &  &  & \\
R2 / R2 adjusted & 0.127 / 0.123 &  &  &  & \\
\bottomrule
\end{tabular}
\caption{Linear regression model results with 95\% CIs. The outcome variable is importance of bias mitigation.}
\end{table}

\begin{table}[H]
\begin{tabular}[t]{lrrrrl}
\toprule
Predictors & Estimate & SE & LL & UL & p\\
\midrule
Intercept & 4.43 &  & 4.31 & 4.56 & <0.001\\
Political orientation & -0.13 &  & -0.20 & -0.06 & <0.001\\
Country (US) & -0.13 &  & -0.33 & 0.08 & 0.216\\
Free speech  & 0.12 &  & 0.05 & 0.19 & 0.001\\
Gender (female) & 0.03 &  & -0.13 & 0.19 & 0.748\\
Education (high) & -0.01 &  & -0.22 & 0.19 & 0.890\\
AI use & 0.29 &  & 0.23 & 0.34 & <0.001\\
Age & 0.01 &  & 0.00 & 0.01 & 0.002\\
Green/Democratic Party & 0.32 &  & 0.09 & 0.55 & 0.006\\
Political orientation X Country (US) & 0.15 &  & 0.05 & 0.24 & 0.003\\
Country (US) X Free speech & -0.15 &  & -0.24 & -0.05 & 0.002\\
Country (US) X Gender (female)  & 0.19 &  & -0.04 & 0.42 & 0.101\\
Country (US) X Education (high) & -0.14 &  & -0.43 & 0.16 & 0.363\\
Country (US) X AI use & -0.14 &  & -0.22 & -0.07 & <0.001\\
Country (US) X Age & -0.00 &  & -0.01 & 0.01 & 0.763\\
Country (US) X Green/Democratic Party & -0.07 &  & -0.40 & 0.26 & 0.684\\
\midrule
Observations & 3556 &  &  &  & \\
R2 / R2 adjusted & 0.052 / 0.048 &  &  &  & \\
\bottomrule
\end{tabular}
\caption{Linear regression model results with 95\% CIs. The outcome variable is importance of showing aspirational version of the world.}
\end{table}

\subsubsection{Single country models}
In this section we report the single models for each country and outcome variable.

\paragraph{Models for the US data only}

\begin{table}[H]
\begin{tabular}[t]{lrrrrl}
\toprule
Predictors & Estimate & SE & LL & UL & p\\
\midrule
Intercept & 5.90 &  & 5.78 & 6.02 & <0.001\\
Political orientation & -0.05 &  & -0.10 & -0.01 & 0.023\\
Free speech & 0.12 &  & 0.08 & 0.17 & <0.001\\
Gender (female) & 0.18 &  & 0.06 & 0.30 & 0.003\\
Education (high) & 0.08 &  & -0.07 & 0.24 & 0.286\\
AI use & -0.01 &  & -0.05 & 0.03 & 0.680\\
Age & 0.01 &  & 0.01 & 0.02 & <0.001\\
Green/Democratic Party & 0.04 &  & -0.13 & 0.21 & 0.631\\
\midrule
Observations & 1756 &  &  &  & \\
R2 / R2 adjusted & 0.061 / 0.057 &  &  &  & \\
\bottomrule
\end{tabular}
\caption{Linear regression model results with 95\% CIs. The outcome variable is importance of accuracy.}
\end{table}

\begin{table}[H]
\begin{tabular}[t]{lrrrrl}
\toprule
Predictors & Estimate & SE & LL & UL & p\\
\midrule
Intercept & 5.24 &  & 5.08 & 5.39 & <0.001\\
Political orientation & -0.04 &  & -0.10 & 0.02 & 0.175\\
Free speech & -0.01 &  & -0.07 & 0.05 & 0.767\\
Gender (female) & 0.69 &  & 0.54 & 0.84 & <0.001\\
Education (high) & -0.09 &  & -0.28 & 0.11 & 0.390\\
AI use & -0.03 &  & -0.08 & 0.02 & 0.224\\
Age & 0.02 &  & 0.02 & 0.03 & <0.001\\
Green/Democratic Party & 0.16 &  & -0.06 & 0.38 & 0.150\\
\midrule
Observations & 1756 &  &  &  & \\
R2 / R2 adjusted & 0.107 / 0.103 &  &  &  & \\
\bottomrule
\end{tabular}
\caption{Linear regression model results with 95\% CIs. The outcome variable is importance of safety.}
\end{table}

\begin{table}[H]
\begin{tabular}[t]{lrrrrl}
\toprule
Predictors & Estimate & SE & LL & UL & p\\
\midrule
Intercept & 5.25 &  & 5.10 & 5.40 & <0.001\\
Political orientation & -0.11 &  & -0.17 & -0.06 & <0.001\\
Free speech & 0.02 &  & -0.04 & 0.07 & 0.525\\
Gender (female) & 0.53 &  & 0.38 & 0.67 & <0.001\\
Education (high) & -0.06 &  & -0.25 & 0.13 & 0.555\\
AI use & 0.03 &  & -0.02 & 0.08 & 0.219\\
Age & 0.02 &  & 0.01 & 0.02 & <0.001\\
Green/Democratic Party & 0.05 &  & -0.17 & 0.26 & 0.674\\
\midrule
Observations & 1756 &  &  &  & \\
R2 / R2 adjusted & 0.078 / 0.075 &  &  &  & \\
\bottomrule
\end{tabular}
\caption{Linear regression model results with 95\% CIs. The outcome variable is importance of bias mitigation.}
\end{table}

\begin{table}[H]
\begin{tabular}[t]{lrrrrl}
\toprule
Predictors & Estimate & SE & LL & UL & p\\
\midrule
Intercept & 4.31 &  & 4.14 & 4.47 & <0.001\\
Political orientation & 0.02 &  & -0.05 & 0.08 & 0.637\\
Free speech & -0.02 &  & -0.09 & 0.04 & 0.512\\
Gender (female) & 0.22 &  & 0.05 & 0.38 & 0.012\\
Education (high) & -0.15 &  & -0.37 & 0.07 & 0.183\\
AI use & 0.15 &  & 0.09 & 0.20 & <0.001\\
Age & 0.01 &  & 0.00 & 0.01 & 0.004\\
Green/Democratic Party & 0.26 &  & 0.01 & 0.50 & 0.039\\
\midrule
Observations & 1756 &  &  &  & \\
R2 / R2 adjusted & 0.026 / 0.022 &  &  &  & \\
\bottomrule
\end{tabular}
\caption{Linear regression model results with 95\% CIs. The outcome variable is importance of showing aspirational version of the world.}
\end{table}

\paragraph{Models for the German data only}

\begin{table}[H]
\begin{tabular}[t]{lrrrrl}
\toprule
Predictors & Estimate & SE & LL & UL & p\\
\midrule
Intercept & 4.94 &  & 4.83 & 5.06 & <0.001\\
Political orientation & -0.12 &  & -0.19 & -0.06 & <0.001\\
Free speech & 0.25 &  & 0.19 & 0.32 & <0.001\\
Gender (female) & 0.17 &  & 0.02 & 0.32 & 0.026\\
Education (high) & 0.43 &  & 0.24 & 0.62 & <0.001\\
AI use & 0.18 &  & 0.13 & 0.23 & <0.001\\
Age & 0.00 &  & -0.00 & 0.01 & 0.363\\
Green/Democratic Party & 0.30 &  & 0.09 & 0.52 & 0.006\\
\midrule
Observations & 1800 &  &  &  & \\
R2 / R2 adjusted & 0.078 / 0.075 &  &  &  & \\
\bottomrule
\end{tabular}
\caption{Linear regression model results with 95\% CIs. The outcome variable is importance of accuracy.}
\end{table}

\begin{table}[H]
\begin{tabular}[t]{lrrrrl}
\toprule
Predictors & Estimate & SE & LL & UL & p\\
\midrule
Intercept & 5.07 &  & 4.95 & 5.20 & <0.001\\
Political orientation & -0.10 &  & -0.18 & -0.03 & 0.005\\
Free speech & 0.21 &  & 0.14 & 0.28 & <0.001\\
Gender (female) & 0.29 &  & 0.13 & 0.45 & <0.001\\
Education (high) & 0.21 &  & 0.01 & 0.41 & 0.041\\
AI use & 0.14 &  & 0.09 & 0.20 & <0.001\\
Age & 0.01 &  & 0.00 & 0.02 & <0.001\\
Green/Democratic Party & 0.31 &  & 0.08 & 0.54 & 0.008\\
\midrule
Observations & 1800 &  &  &  & \\
R2 / R2 adjusted & 0.053 / 0.050 &  &  &  & \\
\bottomrule
\end{tabular}
\caption{Linear regression model results with 95\% CIs. The outcome variable is importance of safety.}
\end{table}

\begin{table}[H]
\begin{tabular}[t]{lrrrrl}
\toprule
Predictors & Estimate & SE & LL & UL & p\\
\midrule
Intercept & 4.67 &  & 4.54 & 4.79 & <0.001\\
Political orientation & -0.16 &  & -0.23 & -0.09 & <0.001\\
Free speech & 0.14 &  & 0.07 & 0.21 & <0.001\\
Gender (female) & 0.11 &  & -0.04 & 0.27 & 0.148\\
Education (high) & 0.12 &  & -0.08 & 0.31 & 0.244\\
AI use & 0.26 &  & 0.20 & 0.31 & <0.001\\
Age & 0.01 &  & 0.00 & 0.02 & <0.001\\
Green/Democratic Party & 0.40 &  & 0.18 & 0.62 & <0.001\\
\midrule
Observations & 1800 &  &  &  & \\
R2 / R2 adjusted & 0.075 / 0.071 &  &  &  & \\
\bottomrule
\end{tabular}
\caption{Linear regression model results with 95\% CIs. The outcome variable is importance of bias mitigation.}
\end{table}

\begin{table}[H]
\begin{tabular}[t]{lrrrrl}
\toprule
Predictors & Estimate & SE & LL & UL & p\\
\midrule
Intercept & 4.43 &  & 4.31 & 4.55 & <0.001\\
Political orientation & -0.13 &  & -0.20 & -0.06 & <0.001\\
Free speech & 0.12 &  & 0.06 & 0.19 & <0.001\\
Gender (female) & 0.03 &  & -0.13 & 0.18 & 0.737\\
Education (high) & -0.01 &  & -0.21 & 0.18 & 0.886\\
AI use & 0.29 &  & 0.24 & 0.34 & <0.001\\
Age & 0.01 &  & 0.00 & 0.01 & 0.001\\
Green/Democratic Party & 0.32 &  & 0.10 & 0.54 & 0.004\\
\midrule
Observations & 1800 &  &  &  & \\
R2 / R2 adjusted & 0.078 / 0.074 &  &  &  & \\
\bottomrule
\end{tabular}
\caption{Linear regression model results with 95\% CIs. The outcome variable is importance of showing aspirational version of the world.}
\end{table}

\subsection{Specification curve analysis}
We additionally assessed with a specification curve analysis (SCA) how robust our findings are. In an SCA, all possible combinations of covariates and the model without any covariates are tested, and the main effects with 95\%-CIs are ordered by size. For our single-variable estimates, we could estimate for each predictor 128 unique combinations (see Figure 4). For each main interaction effect, we could estimate 728 unique combinations with the covariates that were added as interaction with the country or as single models (see Figure5). Overall, the main findings reported in the paper remain robust across all tested specifications. All curves were generated from OLS regression models.

\begin{figure}[!htb]
\centering
\includegraphics[width=\textwidth,height=\textheight,keepaspectratio]{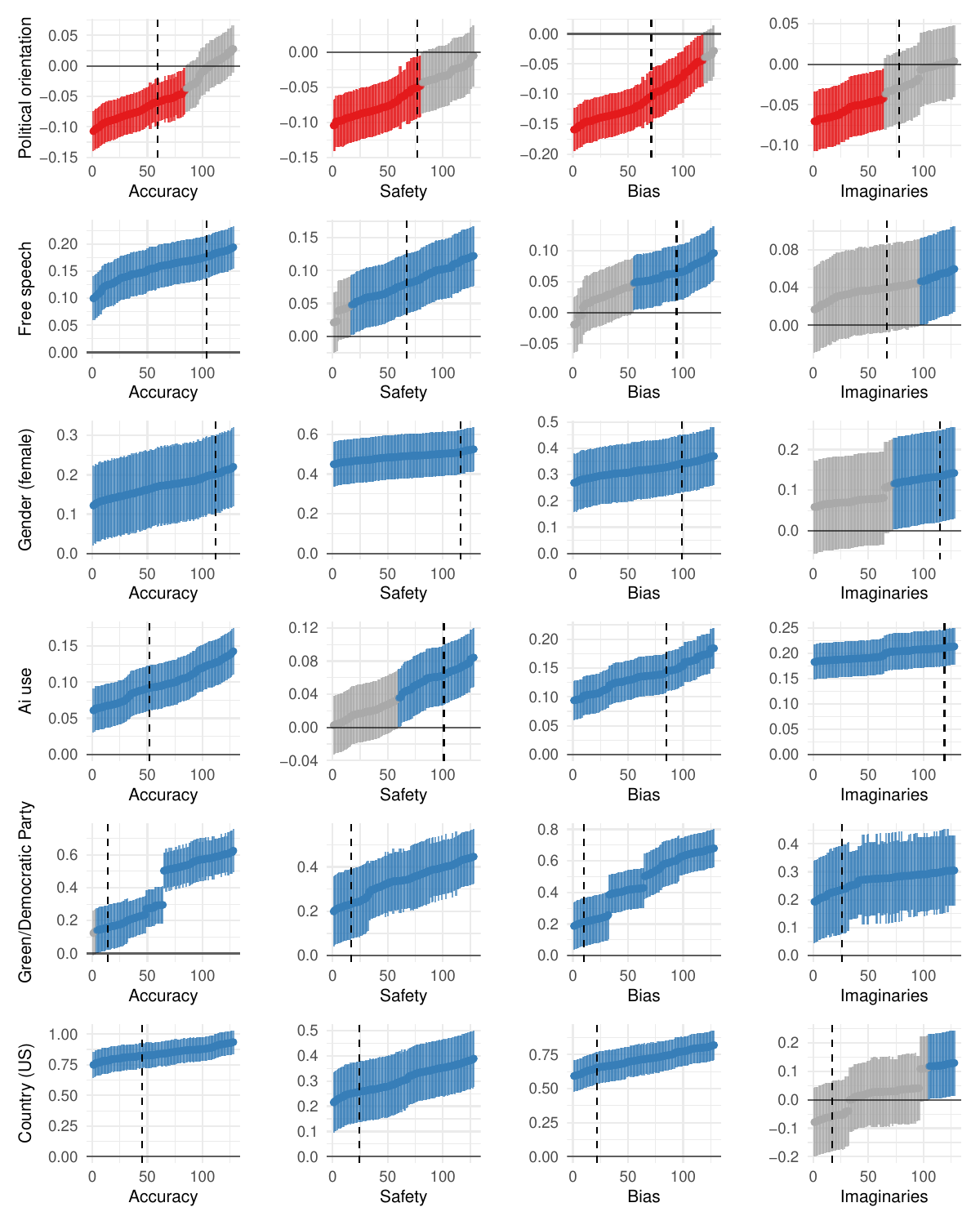}
\caption{Specification curve analysis for all main predictors as single variables without interaction terms. The dashed line indicates the estimate of the model reported in the main paper. A gray line indicates overlap with 0.}
\label{fig:fig1}
\end{figure}

\begin{figure}[!htb]
\centering
\includegraphics[width=\textwidth,height=\textheight,keepaspectratio]{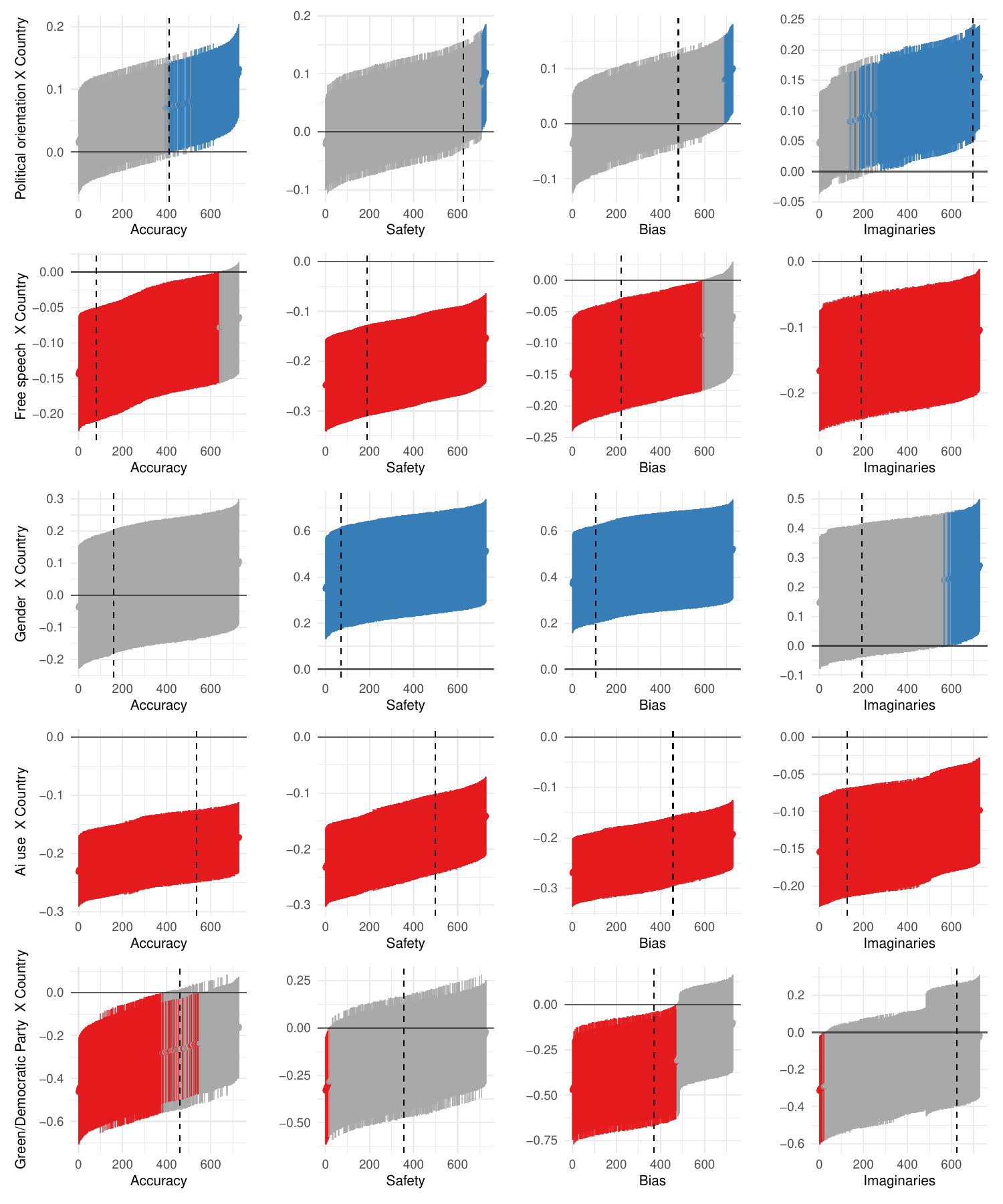}
\caption{Specification curve analysis for all main variables as interaction terms with country. The dashed line indicates the estimate of the model reported in the main paper. A gray line indicates overlap with 0.}
\label{fig:fig2}
\end{figure}

\end{document}